\documentclass[sigconf]{acmart}

\usepackage{graphicx}
\usepackage{subfiles}

\usepackage{todonotes}
\usepackage{booktabs}
\PassOptionsToPackage{hyphens}{url}\usepackage{hyperref}

\usepackage{dsfont}
\usepackage{amsmath}

\usepackage{caption}
\usepackage{subcaption}
\usepackage{xcolor}
\usepackage{listings}

\DeclareFixedFont{\ttb}{T1}{txtt}{bx}{n}{7.8pt} %
\DeclareFixedFont{\ttm}{T1}{txtt}{m}{n}{7.8pt}  %

\definecolor{deepblue}{rgb}{0,0,0.5}
\definecolor{deepred}{rgb}{0.6,0,0}
\definecolor{deepgreen}{rgb}{0,0.5,0}

\newcommand\pythonstyle{\lstset{
language=Python,
basicstyle=\ttm,
morekeywords={self},              %
keywordstyle=\ttb\color{deepblue},
emph={tarexp,ir_measures,component,sklearn,linear_model,LogisticRegression,datasets}, %
emphstyle=\ttb\color{deepgreen},    %
stringstyle=\color{deepred},
frame=tb,                         %
showstringspaces=false
}}
\lstnewenvironment{python}[1][]
{
\pythonstyle
\lstset{#1}
}
{}

\newcommand{\TARexp}{{\texorpdfstring{\texttt{TAR\lowercase{exp}}}{}}}

\AtBeginDocument{%
  \providecommand\BibTeX{{%
    \normalfont B\kern-0.5em{\scshape i\kern-0.25em b}\kern-0.8em\TeX}}}

\copyrightyear{2022}
\acmYear{2022}
\setcopyright{acmlicensed}\acmConference[SIGIR '22]{Proceedings of the 45th International ACM SIGIR Conference on Research and Development in Information Retrieval}{July 11--15, 2022}{Madrid, Spain}
\acmBooktitle{Proceedings of the 45th International ACM SIGIR Conference on Research and Development in Information Retrieval (SIGIR '22), July 11--15, 2022, Madrid, Spain}
\acmPrice{15.00}
\acmDOI{10.1145/3477495.3531663}
\acmISBN{978-1-4503-8732-3/22/07}

\begin{document}

\title[{\TARexp: A Python Framework for Technology-Assisted Review Experiments}]{\TARexp: A Python Framework for \\ Technology-Assisted Review Experiments}

\author{Eugene Yang}
\affiliation{%
  \institution{HLTCOE, Johns Hopkins University}
  \city{Baltimore}
  \state{MD}
  \country{USA}
}
\email{eugene@ir.cs.georgetown.edu}

\author{David D. Lewis}
\affiliation{%
  \institution{Redgrave Data}
  \city{Chantilly}
  \state{VA}
  \country{USA}
}
\email{sigir2022paper@davelewis.com}

\renewcommand{\shortauthors}{Yang and Lewis}

\begin{abstract}
Technology-assisted review~(TAR) is an important industrial application of  information retrieval (IR) and machine learning (ML). While a small TAR research community exists, the complexity of TAR software and workflows is a major barrier to entry.  Drawing on past open source TAR efforts, as well as design patterns from the IR and ML open source software, we present an open source Python framework for conducting experiments on TAR algorithms. Key characteristics of this framework are declarative representations of workflows and experiment plans, the ability for components to play variable numbers of workflow roles, and state maintenance and restart capabilities. Users can draw on reference implementations of standard TAR algorithms while incorporating novel components to explore their research interests. The framework is available at \texttt{\url{https://github.com/eugene-yang/tarexp}}.
\end{abstract}



\maketitle

\begin{figure*}[t]
    \begin{python}
setting = component.combine(component.SklearnRanker(LogisticRegression, solver='liblinear'), 
                            component.PerfectLabeler(), 
                            component.RelevanceSampler(), 
                            component.FixedRoundStoppingRule(max_round=20))()
workflow = tarexp.OnePhaseTARWorkflow(dataset, setting, seed_doc=[1023], batch_size=200, random_seed=123)

recording_metrics = [ir_measures.RPrec, tarexp.OptimisticCost(target_recall=0.8, cost_structure=(25,5,5,1))]
for ledger in workflow:
    print("Round {}: found {} positives in total".format(ledger.n_rounds, ledger.n_pos_annotated)) 
    print("metric:", workflow.getMetrics(recording_metrics))
    \end{python}
\vspace{-1em}
\caption{Sample Python Snippet for Running One-Phase TAR Workflow. Please refer to the online Google Colab Notebook demo page for a full working example.}\label{code:workflow}
\end{figure*}
\section{Introduction}

Technology-assisted review (TAR) is the use of information retrieval (IR) and machine learning (ML) technologies to reduce the cost and increase the effectiveness of manual review of large text collections. Application areas include legal discovery~\cite{baron2016perspectives}, systematic literature review in medicine~\cite{wallace2010semi}, construction of evaluation collections~\cite{hc4}, and responses to sunshine law requests~\cite{mcdonald2020accuracy}. 

Workshops such as DESI\footnote{\url{users.umiacs.umd.edu/~oard/desi7}},  SIRE\footnote{\url{http://users.umiacs.umd.edu/~oard/sire11/}}, LegalAIIA\footnote{\url{https://sites.google.com/view/legalaiia-2021/home}}, and ALTARS\footnote{\url{http://altars2022.dei.unipd.it/}} 
have brought these applications to the awareness of the research community. Shared evaluation efforts such as the TREC Legal Track \cite{treclegal06,treclegal07,treclegal08,treclegal09,treclegal10}, the TREC Total Recall Track~\cite{totalrecall2015,totalrecall2016}, and the CLEF eHealth Technology-Assisted Review Tasks~\cite{clef2017ehealth-tar,clef2018ehealth-tar,clef2019ehealth-tar} have made data sets and formalized evaluation approaches available to researchers.   

However, the inherent complexity of TAR tasks, even when abstracted to research data sets, still imposes a substantial barrier to research. Many research questions in TAR focus on optimizing interactions between cost and effectiveness during an evolving review process. Testing a new TAR approach requires exploring variations of, and capturing rich data from, iterative active learning processes and multiple review stages. Further, the dynamics of these algorithms varies strongly not only across tasks (whether real or simulated) but even across choices of starting conditions and random seeds. Expensive large scale experiments are therefore necessary to derive meaningful generalizations. 
Finally, sample-based effectiveness estimation is itself an object of study in TAR (driven largely by the needs of legal applications) \cite{qbcb-paper}. This raises the stakes for consistency and replicability of evaluation.    

\TARexp~is an open-source Python framework intended to reduce barriers to entry for TAR research. We draw on design patterns from operational TAR software and past open source TAR efforts, as well as those from the broader machine learning and information retrieval open source ecosystem, including \texttt{libact}~\cite{libact} , pyTorch~\cite{pytorch}, pyTerrier~\cite{pyterrier}, ir-datasets~\cite{irds}, and ir-measures~\cite{irm}. 
\TARexp~allows configuration and  dispatching of experimental runs with support for parallel processing, resumption and extension of runs, and reproducibility. It incorporates reference implementations of key components such as supervised learning, active learning, stopping rules, sample-based estimation of effectiveness, and TAR-specific cost visualizations~\cite{cost-structure-paper}. Interfaces in the form of abstract classes for these components aid researchers in implementing and studying their own approaches. The framework is also compatible with Jupyter Notebooks for running exploratory experiments and visualizing results. The framework is available at \texttt{\url{https://github.com/eugene-yang/tarexp}}, and a live demo is available on Google Colab\footnote{\url{https://colab.research.google.com/github/eugene-yang/tarexp/blob/main/examples/exp-demo.ipynb}}.

\section{Background}

Shared software resources in IR research date back to the 1960s~\cite{salton1965smart}. Numerous open source research-oriented retrieval libraries are in use, including Anserini~\cite{yang2017anserini}, pyTerrier~\cite{pyterrier}, Indri~\cite{strohman2005indri}, Galago~\cite{cartright2012galago}, and  Patapsco~\cite{patapsco}.  An even larger ecosystem supports research on ML and natural language processing (NLP), including core IR tasks such as text categorization, summarization, question answering, and recommendation \cite{amiri2019framework, zacharias2018survey, kalyanathaya2019advances}.

However, the needs of TAR research are most similar to those in active learning, an area that has seen less open source activity. We have drawn on ideas from libact~\cite{libact}
\footnote{\url{https://github.com/ntucllab/libact}}
, the most notable open source active learning framework. The structure of \TARexp\ is inspired by its modularization of learning algorithms, sampling strategies, and labelers, but we take a more flexible approach. In particular, we allow components that perform multiple roles in a workflow to support, for example, stopping rules that draw on training sample-based estimates. 

Providing open source code with published research papers is increasingly common, including in several recent TAR studies~\cite{quantstop-paper, li2020stop, yang2019icail}. These reference implementations are useful in replicating particular studies, but have been less useful at supporting TAR research more broadly. 

There have been a few open source efforts focused on broader support for TAR experimentation. The TAR Evaluation Toolkit\footnote{\url{https://cormack.uwaterloo.ca/tar-toolkit/}}  enables simulation of a fixed set of active learning workflows on a labeled version of the Enron collection, and was used in several research studies by the tool's authors \cite{cormack2016engineering, zhang2016sampling, cormack2016scalability}. The Baseline Model Implementation~\cite{cormack2015autonomy} (BMI) is a successor to the TAR Evaluation Toolkit that is wrapped with a VirtualBox virtual machine that provides an interface for users to run TAR interactively. It was used in the TREC Total Recall Tracks as baseline systems\cite{totalrecall2015,totalrecall2016}.  HiCAL\footnote{\url{https://hical.github.io}} embeds BMI in a Django-based framework along with the Indri search engine~\cite{strohman2005indri} and an interface for interactive assessment~\cite{abualsaud2018system}. Components communication through HTTP APIs and so HiCAL has more potential for modification than the previous efforts. It was been used in annotation of the HC4 collections~\cite{hc4}.  

FreeDiscovery\footnote{\url{https://tryfreediscovery.com/}} wraps a REST API around selected scikit-learn IR and ML learning functionality, as well as providing new algorithms for eDiscovery tasks such as email threading and duplication detection.  It does not incorporate support for active learning experiments itself, but has been used as a component in active learning experiments~\cite{yang2017icail}. 

Numerous open source or web-based tools are available for carrying out systematic reviews\footnote{\url{http://systematicreviewtools.com/}}, but most provide little support for algorithmic experimentation. One exception is ASReview\footnote{\url{https://github.com/asreview/asreview}}
, an open source tool implemented in Python and Javascript \cite{van2021open}. It includes a simulation mode that allows running experiments on labeled data sets, and supports user configuration of supervised learning, active learning, and feature extraction methods\footnote{\url{https://asreview.nl/blog/simulation-mode-class-101}}.

\section{Structure of \TARexp}

\begin{figure}
    \centering
    \includegraphics[width=\linewidth]{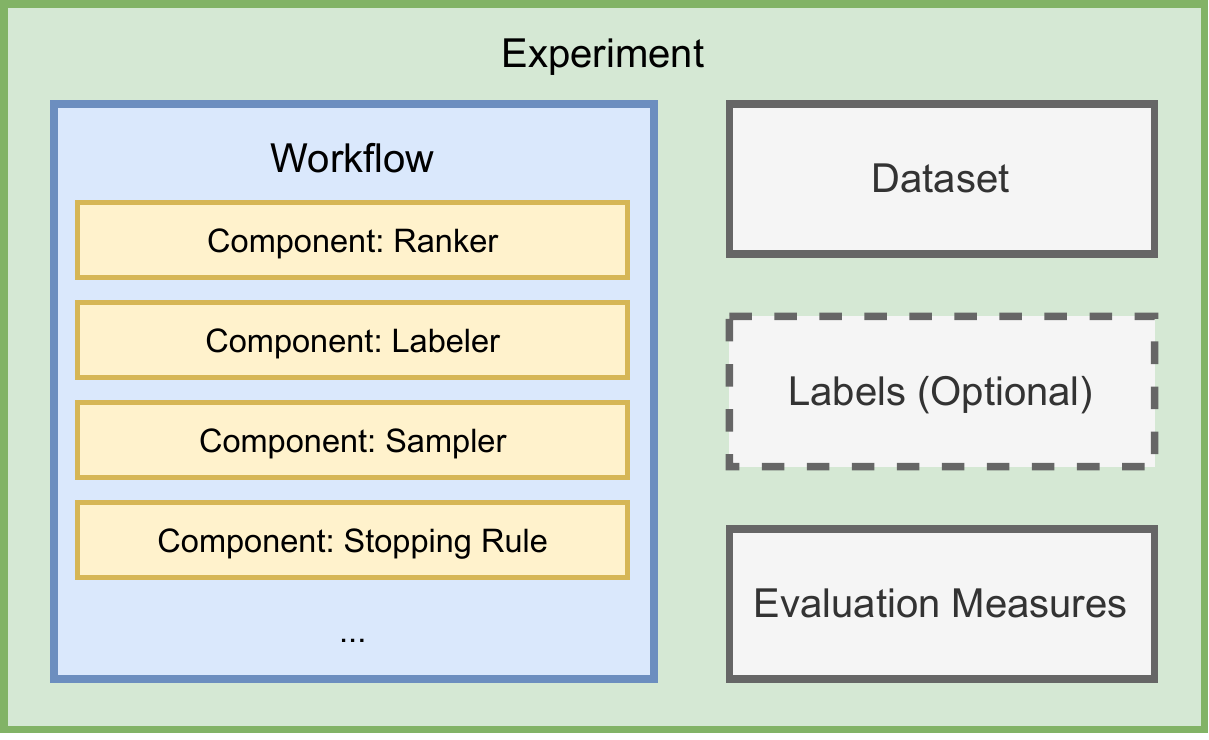}
    \caption{Structure overview of \TARexp. }
    \label{fig:structure}
\end{figure}

A major advance of  \TARexp\ over previous TAR research software is the ability to declaratively specify TAR workflows.  
Users can create components defined using a standard interface and combine them with \TARexp\ components in workflows of their design. This includes incorporating different simulations of human-in-the-loop reviewing, or even embedding in systems using actual human review (though we have not done the latter).

Execution of declaratively specified review workflows is supported by a \textit{workflow}~object (Sec~\ref{sec:structure:workflow}). The changes in the labeling state of the document collection are recorded in the the \textit{ledger}~(Sec~\ref{sec:structure:ledger}). During the iterative process, the workflow reaches out to a set of workflow  \textit{components}~(Sec~\ref{sec:structure:component}), each of which can play one or more roles in the workflow. Finally, an  \textit{experiment}~(Sec~\ref{sec:structure:experiment}) object defines a set of experiments and dispatches them sequentially or in parallel.
Figure~\ref{code:workflow} is a code snippet that demonstrates how each element combines to form a working TAR process, Figure~\ref{fig:structure} is a general overview diagram of \TARexp.  

\subsection{Workflow}\label{sec:structure:workflow}

An object of class \textit{workflow} executes the user's declarative specification of a TAR workflow.  In doing so, it reaches out to \textit{components} for services specified in the declarative specification such as creating training batches, scoring and ranking the collection, and testing for stopping conditions.  

After an optional initial seed round where the user can specify a starting set of labeled training data, the workflow is executed as a sequence of training rounds. Each round consists of selecting a batch of training documents (using a \textit{sampler} object), looking up labels for those documents (using the \textit{labeler} object), training a model and scoring and ranking the collection documents (using the \textit{ranker} object).

\TARexp~supports specifications of both one and two-phase TAR workflows, as described in \citet{cost-structure-paper}. One-phase workflows (\texttt{tarexp.OnePhaseTARWorkflow} in code) can be run for a fixed number of training rounds, or until all documents have been reviewed.  Two-phase reviews also use a stopping rule to determine when to end training, but then follow that by ranking the collection with the final trained model and reviewing to a statistically determined cutoff.

The workflow object maintains only enough state in-memory to work through a training round including the seed for random number generators. Besides the optionally written document scores, the rest of the state of the workflow is recorded in the ledger, which is written to secondary storage at user-configurable intervals. This allows easy restarting of crashed runs with minimal redundant work.

The workflow object is implemented as a Python iterator, allowing procedures defined outside the workflow to execute at each round. The iterator yields a \textit{frozen ledger} (see next Section). The user can define a custom per-round evaluation process or record information for later analysis.

\subsection{Ledger}\label{sec:structure:ledger}

Any aspect of the history of a batch-based workflow can, if necessary, be reproduced from a record of which documents were labeled on which training rounds (including any initial seed round).  The \textit{ledger} object records this state in memory, and writes it to disk at user-specified intervals to enable restarts. 

The persisted ledger for a complete run can be used to execute \TARexp~in \textit{frozen} mode where no batch selection, training, or scoring is done.  Frozen mode supports efficient testing of new components that do not change training or scoring, e.g. non-interventional stopping rules~\cite{qbcb-paper}, effectiveness estimation methods, etc. Evaluating stopping rules for two-phase reviews also requires persisting scores of all documents at the end of each training round, an option the user can specify.

\subsection{Components}\label{sec:structure:component}

\begin{figure*}
    \centering
    \includegraphics[width=\linewidth]{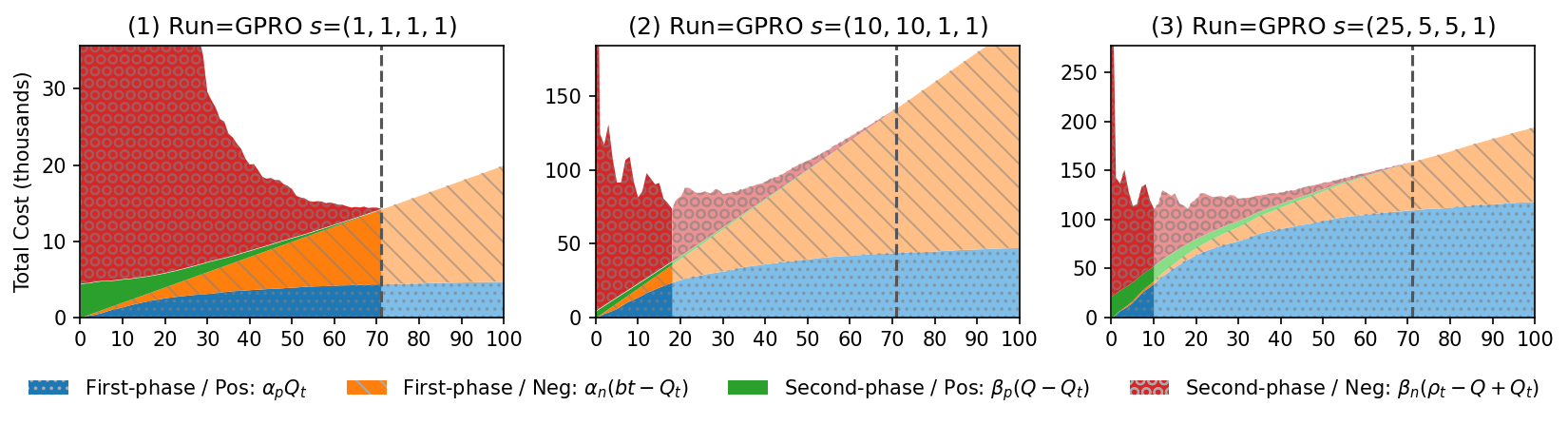}
    \vspace{-2.5em}
    \caption{Cost dynamic graphs on topic GPRO in RCV1-v2 with different cost structured targeting 80\% recall produced by the helper function. 
    The height of the color blocks indicate the cost on each part spent at the respective round. The faded section indicates the rounds that pass the optimal stopping point and the grey vertical dashed line indicates the round where the one phase TAR workflow would reach the recall target.}
    \label{fig:cost}
    \vspace{-1em}
\end{figure*}

\TARexp~implements algorithms via \textit{components}. A component is an object that is declared to serve one or more \textit{roles} in a workflow, e.g. the stopping rule, the training batch sampler, the ranker, or the labeler. Components communicate only through the workflow. The association of components with multiple roles is important when implementing algorithms where, for instance, the stopping rule interacts tightly with a particular batch selection method (e.g. AutoStop~\cite{li2020stop}). The current release of 
\TARexp~defines the interface of multi-role components, but release of the particular multi-role components we have implemented is waiting on a paper under review \cite{betastop-paper}.

\TARexp~supports classification models implemented in Scikit-learn through \texttt{component.SklearnRanker} wrapper. However, any supervised learning model that can produce a score for each document in the collection can be integrated into \TARexp. We have tested an initial implementation of Transformer-based models for TAR \citet{goldilock}, but have not yet integrated this code into the released version of \TARexp.

\TARexp~provides reference implementations of a variety of TAR-specific algorithms, to aid reproducibility and reduce experimenter work. For instance, uncertainty sampling \cite{lewis1994sequential}, relevance feedback \cite{rocchio71relevance}, and simple random sampling batch selection algorithms are provided.

Stopping rules are a particular focus of TAR research. \TARexp~provides implementation of the Knee and Budget Rules~\cite{cormack2016engineering,cormack2015autonomy}, a configurable bath precision rule, the 2399 Rule~\cite{roitblat2007search,quantstop-paper}, fixed numbers of training rounds~\cite{wallace2010semi}, the Quant and QuantCI Rules~\cite{quantstop-paper}, and others. 

A \textit{Labeler} object simulates human review. For most TAR experiments, we assume we simply look up the gold label of each document using \texttt{component.PerfectLabeler}. Random errors can be introduced using \texttt{component.SuccessProbLabeler}. 

\subsection{Evaluation}\label{sec:structure:evaluation}

Consistent implementation of effectiveness metrics, including tricky issues like tiebreaking is critical to TAR experiments. This is true both for evaluation, and because stopping rules may incorporate effectiveness estimates based on small samples. We provide all metrics from the open source package \texttt{ir-measures}\footnote{\url{https://ir-measur.es/en/latest/}} through the 
\texttt{tarexp.Workflow.getMetrics} method. 
Metrics are computed on both the full collection and unreviewed documents to support both finite population and generalization perspectives \cite{yang2019icail}. 

In addition to standard IR metrics, \TARexp~implements \texttt{Optimis-\\ticCost} to support the idealized end-to-end cost analysis for TAR proposed in  \citet{cost-structure-paper}. Such analysis requires specifying a target recall and a cost structure associated with the TAR process (Line 8 in Figure~\ref{code:workflow}). \TARexp also provides helper functions for plotting cost dynamics graphs (Section~\ref{sec:helper:plotting}).

\subsection{Experiments}\label{sec:structure:experiment}

TAR inherits both the large topic-to-topic variability of IR tasks, and the strong dependence on initial conditions and random seeds of active learning processes. Multiple collections, topics, and runs are necessary to reliably demonstrate that one approach dominates another.  Inspired by pyTerrier~\cite{pyterrier}, \TARexp~ supports of a declarative representation for experiments, as shown in this example:

\begin{python}
exp = tarexp.TARExperiment(
    './experiment_output/', 
    random_seed=123, max_round_exec=50,
    metrics=[ir_measures.RPrec, ir_measures.P@10],
    tasks=tarexp.TaskFeeder(
        dataset, 
        labels[['GPRO', 'GOBIT']]),
    components=setting, 
    workflow=tarexp.OnePhaseTARWorkflow,
    batch_size=[200, 50])
results = exp.run(n_processes=2, 
                  n_nodes=2, node_id=0,
                  resume=True, dump_frequency=5)
\end{python}

where an example of defining the variable \texttt{setting} is seen in Figure~1. 
The object \texttt{tarexp.TaskFeeder} creates a stream of tasks, each of which corresponds to a different labeling of the same document collection.  In the example, \texttt{GPRO} and \texttt{GOBIT} are categories in RCV1~\cite{rcv1}. \texttt{components} parameter specify the kind of components for experimenting, which can be either a single one, e.g., the one in Figure~\ref{code:workflow}, or a list of such. Given a task, \texttt{tarexp.TARExperiment} then creates one workflow object for each combination of component, hyperparameter, and random seed choices. The example above will yield four workflows since we have two tasks and two specified batch sizes, and only one alternative was specified for each component.  

We support spawning multiple processes both across machines on a network, and in multiple threads on appropriate hardware.  The method \texttt{.run} dispatches the TAR tasks with runtime settings. In the above example, experiments will run on the first of the two machines with two processes on each, resulting in all four tasks being run simultaneously. The \texttt{.run} method returns the per-round metric values of all the experiment tasks running on the node.

\section{Helper Functions and Integration}

Besides the core functionality of executing experiments, \TARexp~provides several tools to aid in analyzing results. Most are included in our Google Colab notebook. 

\subsection{Experiment Analysis and Visualization}\label{sec:helper:plotting}

The experiment results that \texttt{tarexp.TARExperiment} returns are in generic Python dictionaries. Through \texttt{createDFfromResults}, the results are transformed into Pandas~\cite{pandas} DataFrames for further analysis. The resulting DataFrame contains a multi-level index of the experimental parameters and multi-level columns containing the values of effectiveness and cost metrics. 

We also provide visualization tools to produce cost dynamics graphs, such as Figure~\ref{fig:cost}, through both Python and command-line interfaces. The following is an example command for creating a graph with two runs and three cost structures. 

\begin{python}
python -m tarexp.helper.plotting \
--runs GPRO=location/to/GPRO \ 
       GOBIT=location/to/GOBIT \
--cost_structures 1-1-1-1 10-10-10-1 25-5-5-1 \
--y_thousands --with_hatches
\end{python}

\subsection{Jupyter Notebook and Google Colab}

\begin{figure}
    \centering
    \includegraphics[width=\linewidth]{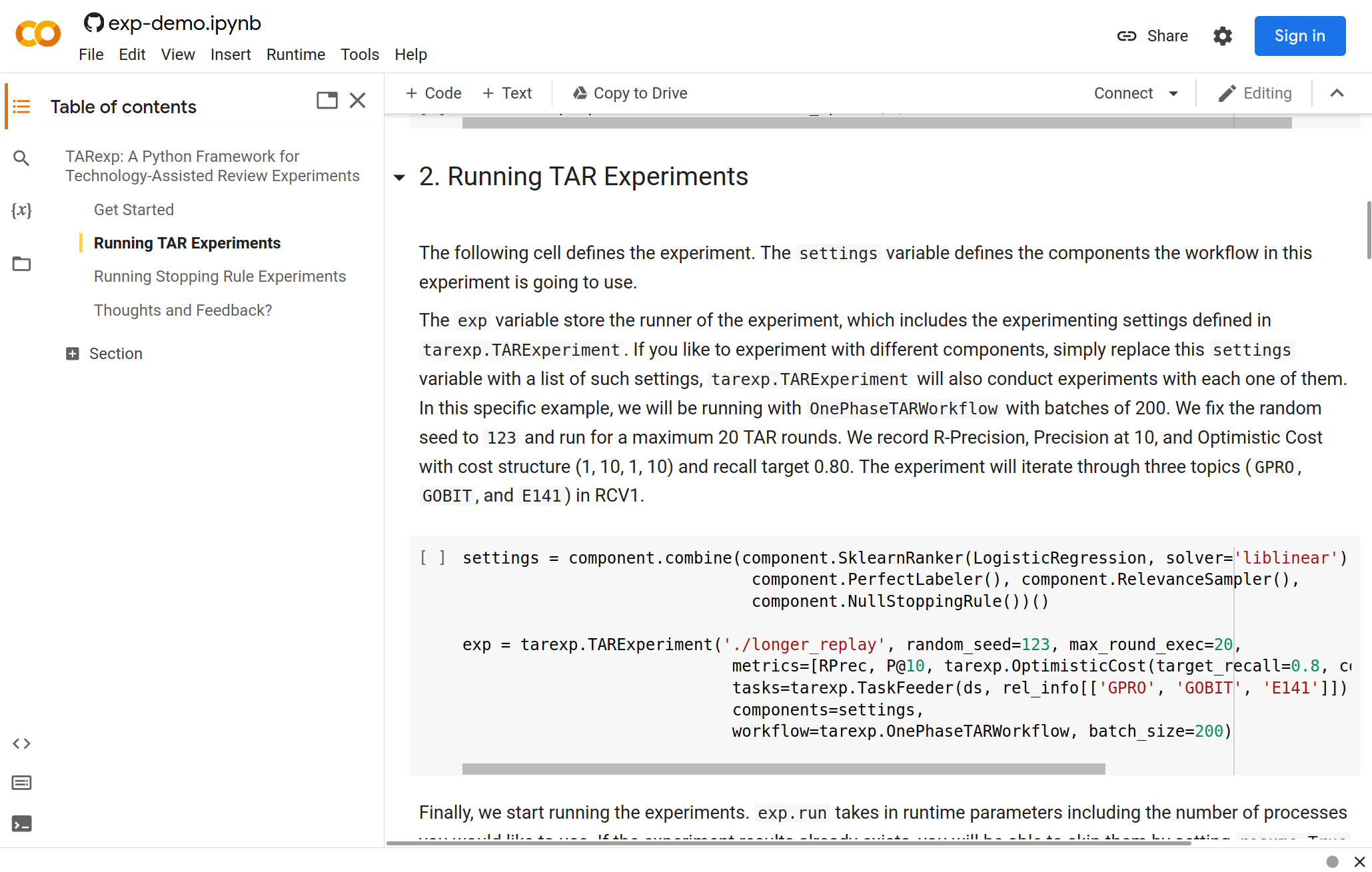}
    \caption{Screenshot of running TAR experiments using \TARexp~ on Google Colab.}
    \label{fig:colab}
\end{figure}

Our framework is fully integrated with Jupyter Notebook~\cite{jupyter} 
, a browser-based tool for running python interactively. Users can also run \TARexp~ on Google Colab\footnote{\url{https://colab.research.google.com/}}, a cloud version of Jupyter Notebook powered by Google, by installing \TARexp~ through Pypi\footnote{\url{https://pypi.org/project/tarexp/}}, the online Python package distribution repository. Figure~\ref{fig:colab} is a screenshot of running TAR experiments on the Google Colab. 
\section{Conclusion}

This paper introduces a new Python framework \TARexp~for conducting TAR experiments, an area of IR involving particularly complex software and workflows.  \TARexp~allows flexible combination of new methods under study with  reference implementations for many commonly used TAR algorithms.  Large scale, deterministically reproducible experiments are supported. We hope that \TARexp~will reduce the barriers to entry for researchers to study the many exciting research problems in TAR, and increase comparability of results.

\begin{acks}
The authors would like to thank Sean MacAvaney and Ophir Frieder for providing valuable feedback on the design of the framework.  
\end{acks}

\bibliographystyle{ACM-Reference-Format}
\bibliography{sample-base}

\end{document}